\documentclass[aps,jap,showpacs,twocolumn,superscriptaddress]{revtex4}

\usepackage{amssymb}
\usepackage{graphicx}
\usepackage{dcolumn}
\usepackage{bm}
\usepackage{color}

\begin{document}

\title{Effect of detuning on the phonon induced dephasing of optically driven InGaAs/GaAs quantum dots}

\author{A.~J.~Ramsay}
\email{a.j.ramsay@shef.ac.uk}
\affiliation{Department of Physics
and Astronomy, University of Sheffield, Sheffield, S3 7RH, United
Kingdom}

\author{T.~M.~Godden}
\affiliation{Department of Physics
and Astronomy, University of Sheffield, Sheffield, S3 7RH, United
Kingdom}

\author{S.~J.~Boyle}
\affiliation{Department of Physics
and Astronomy, University of Sheffield, Sheffield, S3 7RH, United
Kingdom}

\author{E.~M.~Gauger}
\affiliation{Department of Materials, University of Oxford, Oxford OX1 3PH, United Kingdom}

\author{A.~Nazir}
\affiliation{Department of Physics and Astronomy, University College London, London, WC1E 6BT, United Kingdom}

\author{B.~W.~Lovett}
\affiliation{Department of Materials, University of Oxford, Oxford OX1 3PH, United Kingdom}

\author{Achanta~Venu~Gopal}
\affiliation{DCMP \& MS, Tata Institute of Fundamental Research, Mumbai 400 005, India}

\author{A.~M.~Fox}
\affiliation{Department of Physics and Astronomy, University of
Sheffield, Sheffield, S3 7RH, United Kingdom}

\author{M.~S.~Skolnick}
\affiliation{Department of Physics and Astronomy, University of
Sheffield, Sheffield, S3 7RH, United Kingdom}

\date{\today}

\begin{abstract}
Recently, longitudinal acoustic phonons have been identified as the main source of the intensity damping observed in Rabi rotation measurements of the ground-state exciton of a single InAs/GaAs quantum dot. Here we report experiments of intensity damped Rabi rotations in the case of detuned laser pulses, the results have implications for the coherent optical control of both excitons and spins using detuned laser pulses.
\end{abstract}
\maketitle

\section{Introduction}

The crystal ground and neutral exciton states of a single semiconductor quantum dot form a two-level system. As such, resonant excitation of the neutral exciton transition with a laser pulse drives an oscillation in the population inversion known as a Rabi oscillation \cite{Stievater_prl,Zrenner_nat}. A Rabi rotation is typically measured by controlling the pulse-area, the time-integral of the Rabi frequency, via the incident power of the picosecond laser pulse.  In the experiments the Rabi rotation is intensity damped, marking a strong departure from a two-level atom model, provoking a strong debate in the literature concerning the origins of the intensity damping.

In recent experiments, we have identified longitudinal acoustic phonons as the principal source of intensity damping of the ground-state (s-shell) transition of an InAs/GaAs self-assembled dot \cite{Ramsay_prl2010}. We have further demonstrated that the fluctuations in the phonon bath give rise to a temperature and driving field dependent renormalization of the Rabi frequency, and nonmonotonic decay of the Rabi oscillations due to the finite size of the quantum dot \cite{Ramsay_prl2010b}. Here we discuss the underlying physics of the interplay between a driven quantum dot and its phononic environment, and present experiments showing how the LA-phonons modify the absorption spectrum of a picosecond laser pulse.

\section{What is intensity damping?}
\label{sec:intdamp}

\begin{figure}
\begin{center}
\includegraphics[scale=1.5]{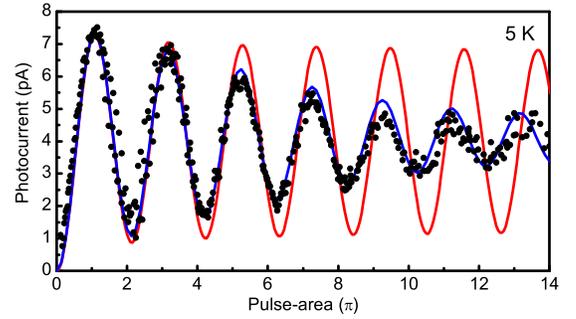}
\end{center}
\caption{($\circ$)A Rabi rotation measurement at $5~\mathrm{K}$. Photocurrent vs square-root of incident power scaled to pulse-area of blue trace. (red) Calculation of Rabi rotation using two-level model with pure dephasing rate of $\Gamma_2^*=0.025~\mathrm{ps^{-1}}$. (blue) Fit using LA-phonon model described in sec. \ref{sec:model}.
}\label{fig:fig1}
\end{figure}

Figure 1 presents a Rabi rotation measurement taken at 5~K. A circularly polarized Gaussian laser pulse excites the neutral exciton transition of a single InAs/GaAs quantum dot on-resonance. The photocurrent which is proportional to the final exciton population is measured as a function of the square-root of the time-averaged incident power $\sqrt{P}$. The x-axis is expressed in terms of the pulse-area $\Theta=\int_{-\infty}^{\infty}\Omega(t')dt'=a\sqrt{P}$, where the Rabi frequency of the pulse is described by $\Omega(t)=\frac{\Theta}{2\tau\sqrt{\pi}}\exp{(-(\frac{t}{2\tau})^2)}$, $\tau=4~\mathrm{ps}$, and $a$ is a measure of the dot-laser coupling. The dot-laser coupling $a$ is deduced from fits to Eqs. (\ref{eq:Blochph_y},\ref{eq:Blochph_z}). Even at this low temperature, the resulting oscillation is damped with increasing power. For full details of the device, and setup see ref. \cite{Boyle_prb}.

To appreciate the relative unimportance of the conventional exponential-type dephasing, the red-trace in fig. 1 presents a calculation of the Rabi rotation using a two-level model with a constant, driving-field independent, rate of dephasing $\Gamma_2^*=0.025~\mathrm{ps^{-1}}$. The red-trace is scaled to best fit the data, and hence the turning points do not occur at integer-$\pi$ pulse-areas. Since the time-duration of the pulse is constant there is a fixed damage to the coherence, and the contrast is lost over the first period of the oscillation, but thereafter is almost constant. By comparison, the data is strongly damped. The damping is nonmonotonic, with the rate of damping decreasing with increasing pulse-area, and the rotation angle is a nonlinear function of the pulse-area, exhibiting a decreases with pulse-area. The blue trace is calculated using the model  in the supplement of ref.  \cite{Ramsay_prl2010}.

\section{Summary of model}

\label{sec:model}

 For the sake of discussion a summary of  the model is presented here. For a full derivation of the model the reader is referred to the supplement of ref. \cite{Ramsay_prl2010}, and refs. \cite{Nazir_prb,Gauger_prb}. The phonon-induced dephasing of an optically driven quantum dot has been the subject of a strong body of theoretical research \cite{Forstner_prl,Machnikowski_prb,Krugel_apb,Krugel_prb,Hohenester_prl,Axt_prb,Hodgson_prb}. A number of alternative approaches of calculating the dephasing have been studied, including a numerically exact approach \cite{Vagov_prl}. Our model treats the quantum dot as an optically driven two-level system coupled to a bath of bosons, in this case LA-phonons. A transformation from the exciton basis to the optically dressed states, the energy eigen-states of the optically driven two-level system, resulting in a stationary frame, where the exciton-phonon interaction is treated as a second-order perturbation. The main approximations are that the phonons are  in thermal equilibrium, and a Born-Markov approximation which assumes that the exciton dynamics are slow compared to the lifetime or memory of the correlation function of the phonon bath $K(t)$. The end result is an intuitive, easy to use model described by a set of three Bloch equations,  with additional Rabi-frequency dependent terms due to the phonon interaction:
\begin{eqnarray}
\dot{s}_x= \Delta s_y -\frac{\Omega^2\Re{[K(\Lambda)]}}{\Lambda^2}s_x -\frac{\Delta\Omega\Re{[K(\Lambda)]}}{\Lambda^2}s_z \nonumber \\ -\frac{\pi\Omega J(\Lambda)}{2\Lambda} \label{eq:Blochph_x}\\
\dot{s}_y= \Omega(1 +\frac{\Im{[K(\Lambda)]}}{\Lambda})s_z -\Delta s_x-\frac{\Omega^2\Re{[K(\Lambda)]}}{\Lambda^2}s_y  \label{eq:Blochph_y}\\
\dot{s}_z= -\Omega s_y \label{eq:Blochph_z}
\end{eqnarray}

The density-matrix is described by the Bloch-vector $\mathbf{s}=(s_x,s_y,s_z)$, where $s_x,s_y$ represent the real and imaginary parts of the excitonic dipole and $s_z$ the population inversion between the exciton and crystal ground-state. The Rabi frequency $\Omega$ and detuning of the laser $\Delta=\omega_X-\omega_{laser}$ result in an  effective Rabi frequency $\Lambda=\sqrt{\Omega^2+\Delta^2}$. The exciton-phonon interaction give rise to additional terms described by:
\begin{eqnarray}
\tilde{K}(t)=\int_0^{\infty} d\omega J(\omega)(2N(\omega)+1)\cos{(\omega t)} \\
K(\Omega)=\int_0^{\infty}dt \tilde{K}(t)e^{i\Omega t} \\
\end{eqnarray}
and the spectral density of the carrier-phonon interaction is
 \begin{equation}
J(\omega) = \sum_{\mathbf{q}}\vert g_{\mathbf{q}}\vert^2\delta(\omega-\omega_{\mathbf{q}}) \label{eq:J}
\end{equation}
where $g_{\mathbf{q}}$ is a generic coupling strength of the exciton to a phonon of wave-vector $\mathbf{q}$. In the case of LA-phonons $g_{\mathbf{q}}$ is  \cite{Krummheuer_prb}
\begin{equation}
g_{\mathbf{q}}=\frac{q(D_e  \mathcal{P}[\psi^{e}(\mathbf{r})] -D_h  \mathcal{P}[\psi^{h}(\mathbf{r})])}{\sqrt{2\mu\hbar\omega_{\mathbf{q}}V}},
\end{equation}
where $\mu$ is the mass density of the host material, $V$ the lattice volume, $D_{e(h)}$ the respective bulk electron (hole) coupling constant, and $\mathcal{P}[\psi^{e(h)}]$ denotes the form factor of the electron (hole) wavefunction.

\section{Interpretation of model: zero detuning $\mathbf{\Delta=0}$}
\label{sec:interpret}

\begin{figure}
\begin{center}
\includegraphics[scale=2.4]{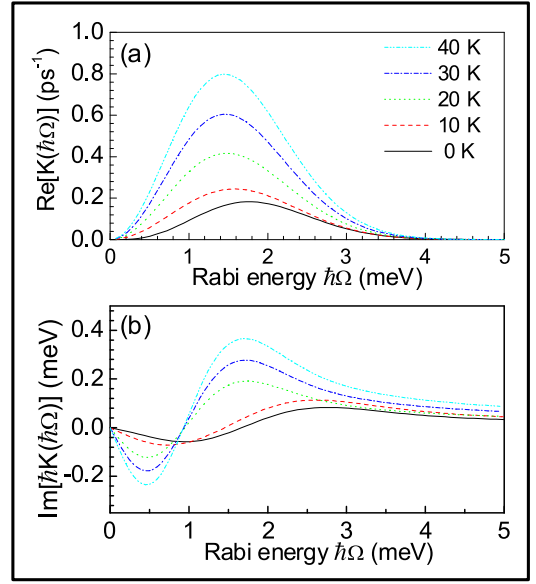}
\end{center}
\caption{Rabi energy and temperature dependence of the exciton-phonon response $K(\Omega)$. (a) The real-part gives a driving-field dependent rate of pure dephasing. (b) The imaginary part gives a renormalization of the Rabi energy.
}\label{fig:fig2}
\end{figure}

Here we consider the case of a Rabi rotation measurement where a single laser pulse excites the neutral exciton transition on-resonance ($\Delta=0$), with the dot initially in the crystal ground-state $\mathbf{s}=(0,0,-1)$. In this case the Bloch equations (\ref{eq:Blochph_x}-\ref{eq:Blochph_z})simplify to:
\begin{eqnarray}
\dot{s}_y= \Omega(1 +\frac{\Im{[K(\Omega)]}}{\Lambda})s_z -\Re{[K(\Omega)]}s_y \label{eq:Blochph_yD0}\\
\dot{s}_z= -\Omega s_y \label{eq:Blochph_zD0}
\end{eqnarray}
\noindent where the phonon interaction is described by a complex response function $K(\Omega)$. The real part describes a Rabi energy dependent rate of pure dephasing, and imaginary term describes a phonon induced shift, or renormalization of the Rabi energy. The real and imaginary parts satisfy a Kramers-Kronig relation.

Figure \ref{fig:fig2} presents temperature dependent calculations of the phonon response function $K(\Omega)$ for the ground-state exciton transition of an InAs/GaAs quantum dot.
To calculate $K(\Omega)$ the spectral density of the exciton-phonon interaction is approximated as $J(\omega)=\alpha\omega^3e^{-\omega^2/\omega_c^2}$, where $\alpha=0.0272~\mathrm{ps^{2}}$, and the cut-off energy $\hbar\omega_c=1.44~\mathrm{meV}$ as measured in ref. \cite{Ramsay_prl2010b}. This
approximation assumes the electron and hole have identical spherical Gaussian wavefunctions, and assumes the phonon modes are those of the bulk host material, in this case GaAs, resulting in a form-factor $\mathcal{P}[\psi^{e,h}(\mathbf{r})]\approx e^{-\omega^2/2\omega_c^2}$ in eq. (\ref{eq:J}).

The real-part $\Re{[K(\Omega)]}=\frac{\pi}{2}\alpha\Omega^3 e^{-\Omega^2/\omega_c^2}\mathrm{coth}(\frac{\hbar\Omega}{2k_BT})$ gives rise to a rate of pure dephasing, and is calculated in fig. \ref{fig:fig2}(a). In a low driving-field regime, at absolute zero, the rate of dephasing scales with the driving-field cubed ($\lim_{\Omega\rightarrow 0, T=0}\Re{K}=\frac{\pi\alpha\Omega^3}{2}$). Whereas at high temperature, in the low driving-field regime, the rate of dephasing is proportional to the square of the driving-field, and the temperature ($\lim_{\Omega\rightarrow 0,T\rightarrow \infty}\Re{[K]}=\frac{\pi\alpha k_BT\Omega^2}{\hbar}\equiv AT\Omega^2$) \cite{Ramsay_prl2010}.  The exciton-phonon interaction has a finite energy bandwidth characterized by a cut-off energy, that corresponds to the energy of a phonon with a wavelength equal to the size of the carrier wavefunction. At high driving-fields exceeding the cut-off energy, the coupling to the phonons with a high energy equal to the Rabi energy is weak, and the phonon-induced dephasing is suppressed. At intermediate driving-fields, this results in a roll-off in the rate of intensity damping at higher pulse-areas \cite{Ramsay_prl2010b}, and in theory the reappearance of high contrast Rabi oscillations at high pulse-area as predicted in \cite{Vagov_prl,Nazir_prb}. Even at absolute zero, the dephasing due to the interplay between the driving field and the LA-phonons can be fierce with a peak rate of dephasing of $\Re{[K_{max}]}=1/(5.6~\mathrm{ps})$.

The imaginary part of the phonon response $\Im{[K]}$ can be calculated numerically as shown in fig. \ref{fig:fig2}(b). This term gives rise to a shift in the rotation speed of the Bloch-vector from $\Omega\rightarrow \Omega_{eff}=\Omega\sqrt{1+\frac{\Im{[K(\Omega)]}}{\Omega}}$, and acts as a rotation that causes an elliptical, rather than a circular orbit of the Bloch-vector, where the ellipse is squeezed along the $s_z$-axis of the Bloch-sphere. The driving-field dependence of $\Im{[K]}$ gives rise to the nonperiodicity of the Rabi rotations observed in ref. \cite{Ramsay_prl2010b}. In a low driving-field regime the effective optical dipole, which we define as $M=\lim_{\Omega\rightarrow 0} \frac{\Omega_{eff}}{\Omega}$ can be calculated as $M=\sqrt{1-\int_0^{\infty}d\omega \frac{J(\omega)}{\omega^2}\mathrm{coth}(\frac{\hbar\omega}{2k_BT})}$. At absolute zero the phonon interaction reduces the effective optical dipole by about 3\%. The effective dipole decreases with temperature, explaining the  increase in the period of the Rabi rotations with temperature observed in ref. \cite{Ramsay_prl2010b}.

A physical picture of the driven phonon dephasing is as follows. The laser drives the charge state of the quantum dot at the Rabi frequency $\Omega$. In turn, the modulation in the charge drives the lattice, resonantly coupling to phonons of the Rabi energy. The frequency dependence of the exciton-phonon coupling, due in part to the density of states, gives rise to a driving-field dependent rate of dephasing and a shift to the rotation angle described by $K(\Omega)$.

A complimentary picture based on the model is that the laser couples the exciton states to form optically dressed states separated by the Rabi energy. Absorption and emission of phonons with an energy equal to the Rabi energy causes relaxation (second-order process) between the dressed states, which is interpreted as a dephasing process in the exciton basis. An experimental visualization of Rabi-split optically dressed states in the time-domain is presented in ref. \cite{Boyle_prl}. This physical picture may be more general. Recent theoretical work suggests that near the band-edge of optical waveguides the radiative decay rate depends on the driving field due to the Rabi side-bands feeling the strong energy gradient in the photon density of states \cite{Ma_prl}. Similar effects have also been suggested in ref. \cite{Melet_prb}.

\section{Influence of detuning}
\label{sec:detune}

\begin{figure}
\begin{center}
\includegraphics[scale=2.5]{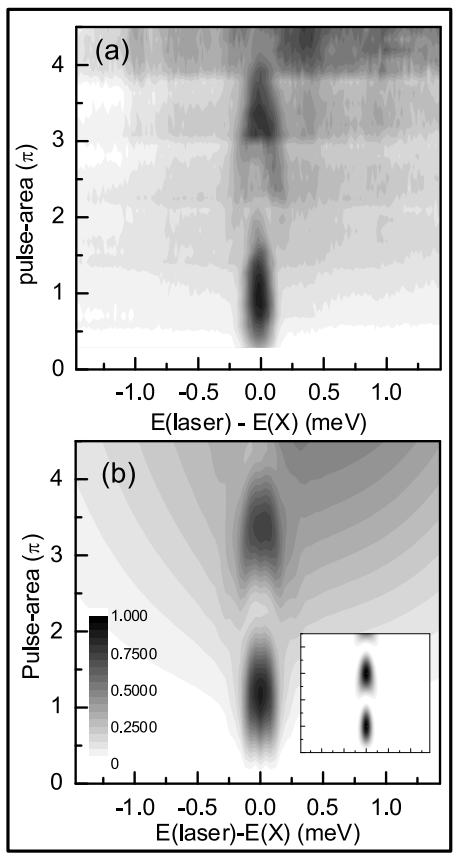}
\end{center}
\caption{(a) Greyscale plot of photocurrent of single quantum dot excited by Gaussian pulse of $\tau=4~\mathrm{ps}$ measured as a function of both laser detuning and pulse-area. The temperature is 15~K, and the reverse bias is 0.6~V. The photocurrent is scaled so that $1\equiv 7.67~\mathrm{pA}$ (b) Calculation of photocurrent using Eqs. (\ref{eq:Blochph_x},\ref{eq:Blochph_y},\ref{eq:Blochph_z}). (inset) Calculation without dephasing.
}\label{fig:fig6}
\end{figure}

Our understanding of the phonon-induced dephasing of an optically driven quantum dot has been experimentally tested in refs. \cite{Ramsay_prl2010,Ramsay_prl2010b}. Here we  examine the role of detuning by studying the effect of the exciton-phonon interaction on the absorption spectrum of a picosecond laser pulse. A number of coherent optical control strategies use detuned \cite{Unold_prl}, or chirped laser pulses \cite{Wu_ArXiv,Simon_ArXiv}. In particular, coherent optical control schemes for a carrier spin usually rely on using detuned laser pulses to minimize the creation of the trion state, see for example \cite{Press_nat,Carter_ArXiv}. Hence, it also important to understand intensity damping in the case of detuned laser pulses.

In the experiment, a single Gaussian laser pulse of $\tau=4~\mathrm{ps}$ excites the quantum dot. A series of photocurrent spectra versus pulse-area are measured by tuning the laser frequency through the ground-state neutral exciton transition. The results are presented in the greyscale plot of fig. \ref{fig:fig6}(a). A peak that closely resembles the spectrum of the laser pulse is modulated by an oscillation in the pulse-area. This peak sits on top of a broad asymmetric feature.

Figure \ref{fig:fig6}(b) presents a calculation of the photocurrent using $\alpha=0.0272~\mathrm{ps^2}$, and $\hbar\omega_c=1.44~\mathrm{meV}$ as determined by the on-resonance experiments \cite{Ramsay_prl2010b}. The pulse-area of fig. \ref{fig:fig6}(a) is scaled to match the model, which reproduces the data quite well. To aid the interpretation, the inset of fig. \ref{fig:fig6}(b) shows a calculation in the case of an ideal two-level system. The broad asymmetric feature arises from photon absorption that is assisted by the emission or absorption of an LA-phonon. At low temperatures, when the LA-phonon population is weak the emission process is stronger resulting in stronger absorption of the blue-detuned laser. This asymmetry is described by the J-term in Eq. \ref{eq:Blochph_x}. A broad shoulder to the quantum dot lineshape due to phonons has also been observed in CW-photoluminescence measurements \cite{Besombes_prb,Favero_prb}.

In the case of the ideal two-level atom, the $3\pi$-peak has a slight arrow-head like shape. The exciton-phonon interaction appears to accentuate arrow-head. For a pulse-area of $2\pi$ the photocurrent spectra develops wings. We note that this effect was reported by Beham {\it et~al} \cite{Beham_physE}, but not explained. As the detuning of the $2\pi$-pulse, and hence the effective Rabi frequency $\Lambda$ is increased the rate of dephasing increases. Consequently, the final state of the Bloch-vector for the detuned pulse resides closer to the center of the Bloch-sphere than for the on-resonant laser pulse, resulting in higher absorption for the detuned pulse and the formation of the wings. The absorption spectra of a quantum dot excited by a rectangular spectrum laser pulse can also exhibit side-peaks corresponding to the edges of the pulse spectrum \cite{Ramsay_prb}, but this is not a factor here.

In this section, we have measured the quantum dot absorption spectrum of a Gaussian laser pulse for various pulse-areas. In addition to the intensity damping of the Rabi oscillation, the exciton-phonon interaction gives rise to a broad, asymmetric lineshape due to phonon-assisted absorption. The absorption spectra is well-described by the model. Based on the model, for small detuning the increase in the effective Rabi frequency enhances the excitation dephasing. To achieve high fidelity operations, the red-detuning of the laser pulse needs to exceed the cut-off energy.

\section{Evidence for phonon-mediated excitation induced dephasing in previous experiments}

The strength of the intensity damping is determined by the $\alpha$-parameter which is a bulk property of the host material. Therefore the strength of the intensity damping for all quantum dots embedded in GaAs should be $\alpha_{GaAs}=0.0272~\mathrm{ps^2}$, as measured in refs. \cite{Ramsay_prl2010,Ramsay_prl2010b}. In this section, we test this hypothesis by comparing the model to other experimental reports of excitation induced dephasing.

In ref. \cite{Stufler_prbR}, Stufler {\it et~al} report a high quality Rabi rotation of the ground-state neutral exciton transition using photocurrent detection. We note that the model can reproduce the intensity damping using  the experimental parameters given in ref. \cite{Stufler_prbR}, and $\alpha=0.0272~\mathrm{ps^2}$. The cut-off energy was used as the only fitting parameter and a value of $\hbar\omega_c=1.67~\mathrm{meV}$ was obtained. This corresponds to an exciton probability distribution with a FWHM of 4.7~nm, which is reasonable for an InAs/GaAs quantum dot.

Off-resonant excitation of wetting layer states has also been suggested as a possible mechanism for the intensity damping \cite{VillasBoas_prl}. In ref. \cite{VillasBoas_prl} the argument is supported by a fit to the data of Zrenner {\it et~al} \cite{Zrenner_nat}. We note however that the strength of the damping used is consistent with LA-phonons at a temperature of 4~K. In ref. \cite{Wang_prb}, the intensity damping of Rabi rotations of p-shell excitons was described  using a rate of dephasing $\Gamma_2=w\left(\frac{\Theta}{\pi\tau_p}\right)^2$ where $\tau_p$ is the intensity FWHM of the sech-pulse used in the experiments, and $c=0.135~\mathrm{ps}$. However, similar results can also be obtained for LA-phonons at a temperature of 5~K, and a cut-off energy of $\sim 1.3~\mathrm{meV}$, suggesting that LA-phonons make a significant contribution to the intensity damping even for p-shell excitons that are energetically close to the wetting layer states.

In experiments to observe an Autler-Townes doublet \cite{Jundt_prl}, a 10\% increase in the exciton linewidth for Rabi energies of up to $120~\mathrm{\mu eV}$ was reported. At 4.2~K, a Rabi energy of $120~\mathrm{\mu eV}$ would give rise to a rate of excitation induced dephasing of $0.5~\mathrm{\mu eV}$, compared to the $5~\mathrm{\mu eV}$ linewidth of the exciton transition. In measurements of a Rabi oscillation driven by a continuous wave laser with Rabi energies of $< 40~\mathrm{\mu eV}$, Flagg {\it et~al} \cite{Flagg_nphys} reported no evidence for excitation induced dephasing. Based on our model, the expected excitation induced dephasing due to phonons would be 56~neV, which was not resolvable in those experiments.

To conclude this section, intensity damping due to LA-phonons is an intrinsic dephasing mechanism. The $\alpha$-parameter that defines the strength of the damping depends on the bulk acoustic properties of the host material. We have explored this by comparing the LA-phonon model to a number of previously reported experiments, and find that the LA-phonon model is consistent with many key observations of intensity dependent dephasing under resonant excitation conditions.

\section{Outlook}

For a solid-state qubit coupled to a phononic environment the rate of dephasing is not characterized by a static number, but depends strongly on the driving field. Achieving high-fidelity quantum logic operations will not be achieved by simply using faster laser pulses, but by optimizing the coherent control scheme to minimize dephasing based on an understanding of a driven qubit coupled to a bath. Whilst the experiments reported here, and in refs. \cite{Ramsay_prl2010,Ramsay_prl2010b} provide strong support for the existing theories, several important aspects remain untested. For example, if the driving-field is strong enough, will the intensity damped Rabi rotation reappear \cite{Vagov_prl,Nazir_prb}? Also, can one observe consequences of the phonon bath memory \cite{Krugel_prb}? Conversely, can the phonon bath be used as a resource? For example, could a laser driven quantum dot be used as a Rabi frequency tuneable point source of LA-phonons; as a nanoscale heat pump \cite{Gauger_prb2010}, or for state inversion \cite{Stace_prl}?

\section{Acknowledgements}

The authors thank the EPSRC (UK) EP/G001642, the QIPIRC UK, and UK-India Education Research Initiative for financial support. AN is supported by the EPSRC, and BWL by the Royal Society. We thank H.~Y.~Liu and M.~Hopkinson for sample growth. AJR thanks the Royal Society for a travel grant to attend ICPS.


\end{document}